\documentclass[lettersize,journal]{IEEEtran}
\usepackage{amsmath,amsfonts}
\usepackage{bibentry}  
\usepackage{algorithmic}
\usepackage{algorithm}
\usepackage{xcolor}
\usepackage{array}
\usepackage[caption=false,font=normalsize,labelfont=sf,textfont=sf]{subfig}
\usepackage{textcomp}
\usepackage{stfloats}
\usepackage{url}
\usepackage{verbatim}
\usepackage{graphicx}
\usepackage{cite}
\begin{document}

\title{Polarization Modulation in Quantum-Dot Spin-VCSELs for Ultrafast Data Transmission}

\author{Christos Tselios,
        Panagiotis Georgiou,
        Christina (Tanya) Politi,
        Antonio Hurtado,
        Dimitris Alexandropoulos
  \thanks{Christos Tselios and Christina (Tanya) Politi are with Department of Electrical and Computer Engineering, University of Peloponnese, Patra, Achaia, 22100, Greece.}
  \thanks{Panagiotis Georgiou and Dimitris Alexandropoulos are with the Department of Materials Science, University of Patras, Rion, Achaia 26504 Greece.}
  \thanks{Antonio Hurtado is with the Institute of Photonics, Department of Physics, University of Strathclyde, Technology and Innovation Centre, Glasgow G1 1RD, U.K..}
  \thanks{Antonio Hurtado acknowledges funding support by the UKRI Turing AI Acceleration Fellowships Programme (EP/V025198/1).}}



\maketitle

\begin{abstract}
Spin-Vertical Cavity Surface Emitting Lasers (spin-VCSELs) are undergoing increasing research effort for new paradigms in high-speed optical communications and photon-enabled computing. To date research in spin-VCSELs has mostly focused on Quantum-Well (QW) devices. However, novel Quantum-Dot (QD) spin-VCSELs, offer enhanced parameter controls permitting the effective, dynamical and ultrafast manipulation of their light emission’s polarization. In the present contribution we investigate in detail the operation of QD spin-VCSELs subject to polarization modulation for their use as ultrafast light sources in optical communication systems. We reveal that QD spin-VCSELs outperform their QW counterparts in terms of modulation efficiency, yielding a nearly two- fold improvement. We also analyse the impact of key device parameters in QD spin-VCSELs (e.g. photon decay rate and intra-dot relaxation rate) on the large signal modulation performance with regard to optical modulation amplitude and eye-diagram opening penalty. We show that in addition to exhibiting enhanced polarization modulation performance for data rates up to $250Gb/s$, QD spin-VCSELs enable operation in dual (ground and excited state) emission thus allowing future exciting routes for multiplexing of information in optical communication links.

\end{abstract}


\section{Introduction}
\IEEEPARstart{S}{pintronic} devices like, spin-polarized Light-Emitting Diodes (LEDs) \cite{holub_spin-polarized_2007}, spin-Vertical Cavity Surface Emitting Lasers (VCSELs) \cite{gerhardt2012spin,lindemann_ultrafast_2019,yokota_spin-injected_2021} and bimodal QD micropillar cavities \cite{heermeier2022spin} are being widely investigated for next-generation optical communication systems. In these devices the radiative recombination of spin-polarized carriers leads to luminescence exhibiting circular polarized emission. There are several advantages of spin devices that have fuelled recent research. These include reduced lasing threshold \cite{junior2015toward} compared to conventional lasers, spin amplification \cite{cemlyn2018near,cemlyn2019polarization} and polarization resolved output \cite{frougier_control_2013,schires_optically-pumped_2012,alharthi_1300_2015}. Spin lasers, including spin VCSELs, exhibit ultrafast dynamics \cite{junior2015toward,lindemann_frequency_2016} due to the spin coupling between electrons and photons in spin-VCSELs thus overcoming the limitation of relaxation oscillation frequency \cite{torre_high_2017} commonly faced by conventional non-spin lasers.     

Harkhoe \textit{et al.} have used polarization modulation in spin-VCSELs to improve the processing speed of a photonic reservoir computing \cite{harkhoe2021neuro}. Yokota \textit{et al.} have used spin-VCSELs as Analog Radio-over-Fiber transmitters for networks beyond $5G$ \cite{yokota_spin-injected_2021}. Lindemann \textit{et al.} demonstrated the potential to enhance the modulation bandwidth of optical communication systems up to $240$ $GHz$ and single channel data transmission rate of $240$ $Gb/s$ \cite{lindemann_ultrafast_2019}. The latter is nearly an order of magnitude higher than what is achieved with conventional directly modulated semiconductor lasers. Recently, Huang \textit{et al.} explored high-frequency polarization modulation in optically pumped spin-VCSELs over a broad range of basic device properties and operating conditions \cite{huang2021optically}.

These aforementioned demonstrations employed Quantum-Well (QW) spin-VCSELs. However, Quantum Dots (QDs) hold much promise for spin devices. Huang \textit{et al.} \cite{huang_room-temperature_2021} demonstrated electron spin polarization exceeding 90$\%$ at room temperature by remote spin filtering of InAs QD electrons without a magnetic field. Giba \textit{et al.} \cite{giba_spin_2020} used QD as gain material in \textit{p-}doped spin-polarized Light-Emitting Diodes (LED) to show a two fold improvement in the degree of circular polarization compared to QW spin-LEDs at zero magnetic field. The spin optoelectronics literature has been recently enriched with few accounts on the potential of QD spin-VCSELs for optical communication and processing applications \cite{huang_high-speed_2021, skontranis2022time}. Huang and co-workers \cite{huang_high-speed_2021} numerically demonstrated a high-speed secure key distribution (SKD) that is enabled by the tunability of the pump polarization ellipticity in QD spin-polarized VCSELs. In \cite{skontranis2022time} Skontranis \textit{et al.} exploited emission from two discrete wavebands and two polarization states to enhance computational efficiency of a dual-state QD spin-VCSEL based Reservoir Computing (RC). The broad spectral bandwidth of QD structure can be deployed in wavelength-division multiplexing systems (WDM).


Motivated by these, we study theoretically  in detail the polarization modulation in QD spin-VCSELs for ultrafast optical communication applications. Our approach demonstrates better performance of QD spin-VCSELs over their QW counterparts in terms of modulation efficiency. The gauge of evaluation are the eye diagrams and the Eye Opening Penalty (EOP). The latter reveals the quality of the dynamic performance of optically modulated signals. To analyse the modulation efficiency in QD spin-VCSELs, we investigate the amplitude of the optical modulation as a function of the photon decay rate and intradot relaxation rate. We use this technique to elucidate the effect of these key parameters on the evolution of the output polarization ellipticity of QD spin-VCSELs. Our investigations reveal ultrafast modulation speeds of $250$ $Gb/s$ (and possibly higher) with QD-spin VCSELs with enhanced modulation efficiency when compared with their QW counterpart. We also show that QD spin-VCSELs yield ultrafast operation from both their ground states (GS) and excited states (ES); thus offering great prospects for ultrafast dual wavelength laser modules.
\section{Theory}

The model used here is an elaboration of the  SFM coupled rate equations \cite{georgiou_effect_2021} for the case of dual-state QD spin-VCSELs that accounts for upwards transitions. It is experimentally demonstrated that carrier escape from the excited and ground levels into the wetting layer has non-negligible effects on the carrier density and spin polarization in both levels \cite{huang_room-temperature_2021}.  The modified rate equations read:
\begin{equation}\label{n_wl}
\begin{split}
\frac{dn_{WL}^\pm}{dt}=\gamma_{n}J^{\pm}-\gamma_{0}n_{WL}^\pm(1-f_{ES}^{\pm})-\gamma_{n}n_{WL}^\pm+\\
\noindent{+4\gamma_{esc}f_{ES}^\pm\mp\gamma_{j}(n_{WL}^{+}-n_{WL}^{-})}
\end{split}
\end{equation}
\begin{equation}\label{n_es}
    \begin{split}
\frac{df_{ES}^\pm}{dt}&=\frac{1}{4}\gamma_{0}n_{WL}^\pm(1-f_{ES}^{\pm})-\gamma_{esc}f_{ES}^\pm-\gamma_{n}f_{ES}^\pm-\\&-\gamma_{21}f_{ES}^{\pm}(1-f_{GS}^\pm)+\frac{1}{2}\gamma_{12}f_{GS}^\pm(1-f_{ES}^{\pm})-\\&-\frac{1}{4}\gamma_{n}(2f_{ES}^\pm-1)|E_{s_{ES}}^{\pm}|^2\mp\gamma_{j}(f_{ES}^{+}-f_{ES}^{-})
\end{split}   
\end{equation}
\begin{equation}\label{n_gs}
\centering
    \begin{split}
\frac{df_{GS}^\pm}{dt}&=2\gamma_{21}f_{ES}^\pm(1-f_{GS}^\pm)-\gamma_{12}f_{GS}^\pm(1-f_{ES}^\pm)-\gamma_{n}f_{GS}^\pm-\\&-\frac{1}{2}\gamma_{n}(2f_{GS}^\pm-1)|E_{s_{GS}}^\pm|^2-\gamma_{j}(f_{GS}^{+}-f_{GS}^{-})
\end{split}    
\end{equation}
\begin{equation}\label{E_s_GS}
\frac{dE_{GS}^\pm}{dt}=k[h_{1}(2f_{GS}^\pm-1)](1+i\alpha)E_{GS}^\pm-(\gamma_{a}+i\gamma_{p})E_{GS}^\mp
\end{equation}
\begin{equation}\label{E_s_ES}
\frac{dE_{ES}^\pm}{dt}=k[h_{2}(2f_{ES}^\pm-1)](1+i\alpha)E_{ES}^\pm-(\gamma_{a}+i\gamma_{p})E_{ES}^\mp
\end{equation}

The dynamical variables $n_{WL}$ and $n_{ES}$ ($n_{GS}$) denote normalized conduction band carrier concentrations of Wetting Layer (WL) and ES (GS) level, respectively. Here, superscripts $+$ and $-$ correspond to the Right Circular Polarized (RCP) and Left Circular Polarized (LCP) components of the emitted light. The SFM parameters for QD spin-VCSEL are defined as follows: $k$ is the photon decay rate, $\alpha$ is the linewidth enhancement factor, $h_{1}$ is the normalized gain coefficient for GS transitions, $h_{2}$ is the normalized gain coefficient for ES transitions, $\gamma_{n}$ is the carrier recombination rate, $\gamma_{21}$ is the intradot relaxation rate of spin polarized carriers relax from ES level at the spin-up(down) GS level, $\gamma_{0}$ is the capture rate from WL into ES level, $\gamma_{j}$ is the spin relaxation rate at ES,GS and WL,  $\gamma_{p}$ is the birefringence rate and $\gamma_{a}$ is the dichroism rate. Some carriers can be excited thermally from ES to WL with escape rate $\gamma_{esc}$ and from GS to ES with escape rate $\gamma_{12}$. These relations read as follows: $\gamma_{12}$ = $\gamma_{21}$$e^{-\frac{\Delta{E_{ES,GS}}}{k_{B}T}}$ and $\gamma_{esc}$ = $\gamma_{0}$$e^{-\frac{\Delta{E_{WL,ES}}}{k_{B}T}}$, where $\Delta{E_{ES,GS}}$ ($\Delta{E_{WL,ES}}$) is the energy of carrier excitation from GS (WL) to ES (WL) and $k_{B}T$ is the product of Boltzmann constant $k_{B}$ and the temperature $T$.

Lasing occurs via transitions from the ES or GS, to the valence band (VB) emitting right $(E_{ES}^+,E_{GS}^+)$ and left $(E_{ES}^-,E_{GS}^-)$ circularly polarized electric fields at two distinct wavelengths. The carrier injection at ES and GS threshold is respectively denoted by $I_{ES,th}^\pm$  and $I_{GS,th}^\pm$.

Polarization control can be realized by means of the pump parameters, namely normalized pump stength $J$, which is equal to $1$ at threshold, and pump ellipticity $P$. The total pump strength is defined by $J=J_{+}+J_{-}$, where $J_{+}$ and  $J_{-}$ are RCP and LCP normalized pump components and the pump polarization ellipticity by: 
\begin{equation}\label{pump_ellipticity}
 P=(J_{+}-J_{-})/(J_{+}+J_{-})   
\end{equation}
When, $P=0$ is applied there is no preference on the spin polarization of the injected carriers $(P=0 \rightarrow J_{+}=J_{-})$ and spin-VCSELs are reduced to conventional VCSELs. The QD spin-VCSEL output is expressed in terms of circularly polarized intensities $I_{ES}^{+}=|E_{ES}^{+}|^{2},I_{ES}^{-}=|E_{ES}^{-}|^{2}$ and the polarization ellipticity $\epsilon$ is defined as:
\begin{equation}\label{output_ellipticity}
\epsilon_{ES}=(I_{ES}^{+}-I_{ES}^{-})/(I_{ES}^{+}+I_{ES}^{-})    
\end{equation}
describing almost circular polarized emission when $\epsilon_{ES} \rightarrow 1$ and linear emission in the case of $\epsilon_{ES} \rightarrow 0$. The same applies for $I_{GS}^{+}$, $I_{GS}^{-}$ and $\epsilon_{GS}$. Throughout this paper we use the following parameter values: $k_{B}T=0.025$ $eV$, $\Delta{E_{ES,GS}}=0.05$ $eV$, $\Delta{E_{WL,ES}}=0.1$ $eV$, $\alpha=3$, $h_{1}=1.995$, $J=100$, $\gamma_{p}=500$ ${ns^{-1}}$, $\gamma_{n}=1$ ${ns^{-1}}$, $\gamma_{0}=$ $800{ns^{-1}}$ and $\gamma_{a}=0$ ${ns^{-1}}$. Higher values of $J$ can improve polarization modulation performance (see Appendix II), however, there is a trade-off as growing $J$ will increase the energy consumption and heating levels in the device.

\section{High speed polarization modulation in dual state QD spin-VCSEL}

\begin{figure}[t]
\centering
\includegraphics[width=\columnwidth]{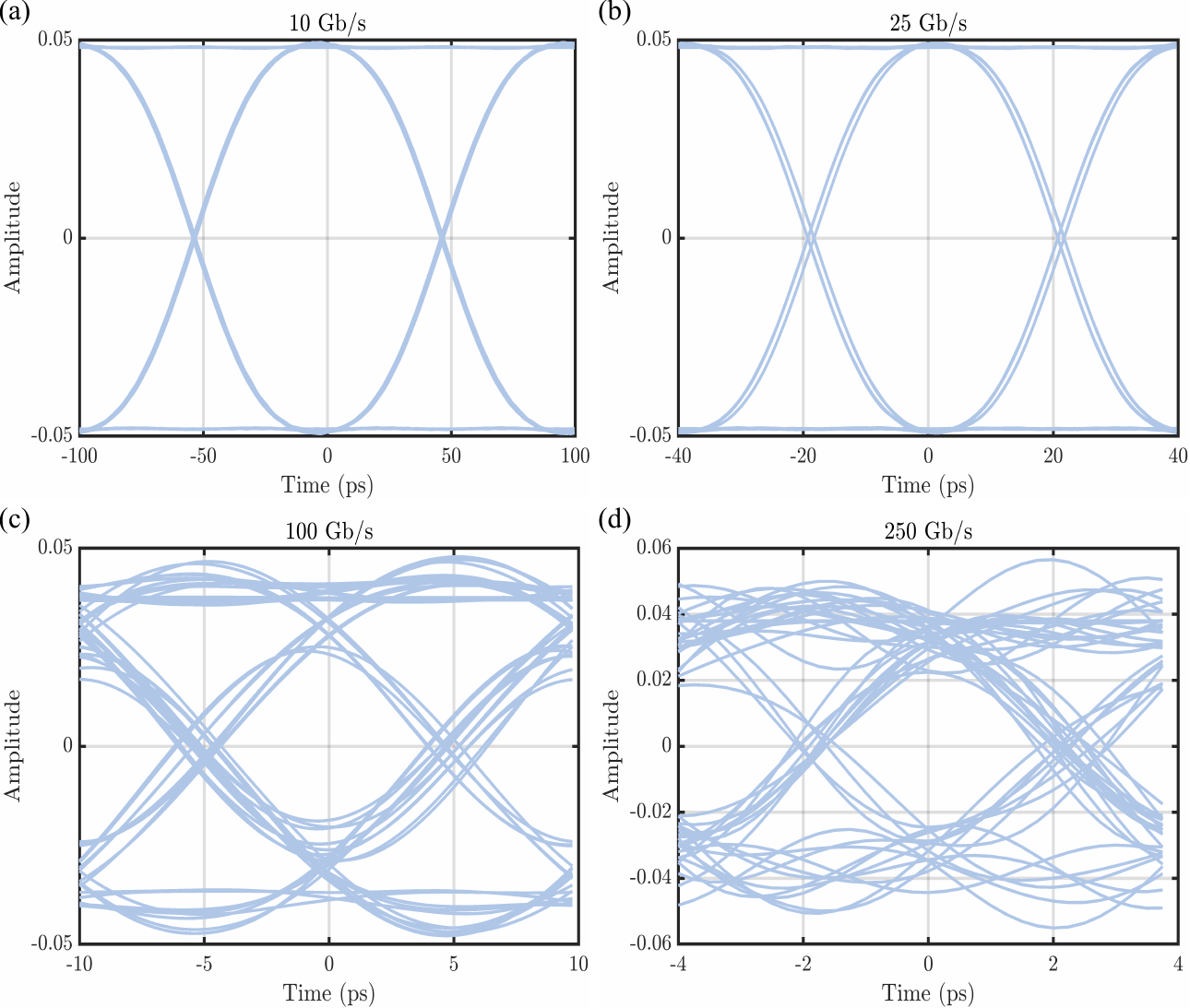}
\caption{2-bit-long eye diagrams of the ES of the QD spin-VCSEL subject to polarization modulation for bit sequences of $10$ $Gb/s$ $(a)$, $25$ $Gb/s$ $(b)$, $100$ $Gb/s$ $(c)$ and $250$ $Gb/s$ $(d)$. For all panels: $\gamma_{j}=250$ $ns^{-1}$, $\gamma_{p}=500$ $ns^{-1}$ and $k=250$ $ns^{-1}$.}
\label{fig:eyes}
\end{figure}

Optical pumping using circularly polarized light can generate spin-polarized electron population under optical selection rules. Spin-up electrons are related to LCP emission, while spin-down electrons result in RCP emission \cite{gerhardt2012spin}. 
Based on this, we can modulate the QD spin-VCSEL's output ellipticity by dynamically controlling the input ellipticity with fast time-varying optical spin injection pulses. To do this, we implement a Non-Return-to-Zero (NRZ) polarization modulation scheme to the QD spin-VCSEL. 
Bits '1' and '0' are defined as RCP and LCP pumping respectively \cite{lindemann_ultrafast_2019}. The bit stream ($127$ pseudorandom bits) is filtered by a raised-cosine filter in order to minimize the intersymbol interference \cite{wasner2015digital}. The corresponding $\epsilon$,  which is calculated by means of polarized intensities (see equation \ref{output_ellipticity}), contains the information encoded into $P$.
\begin{figure*}[!ht]
\centering
\includegraphics[width=\linewidth]{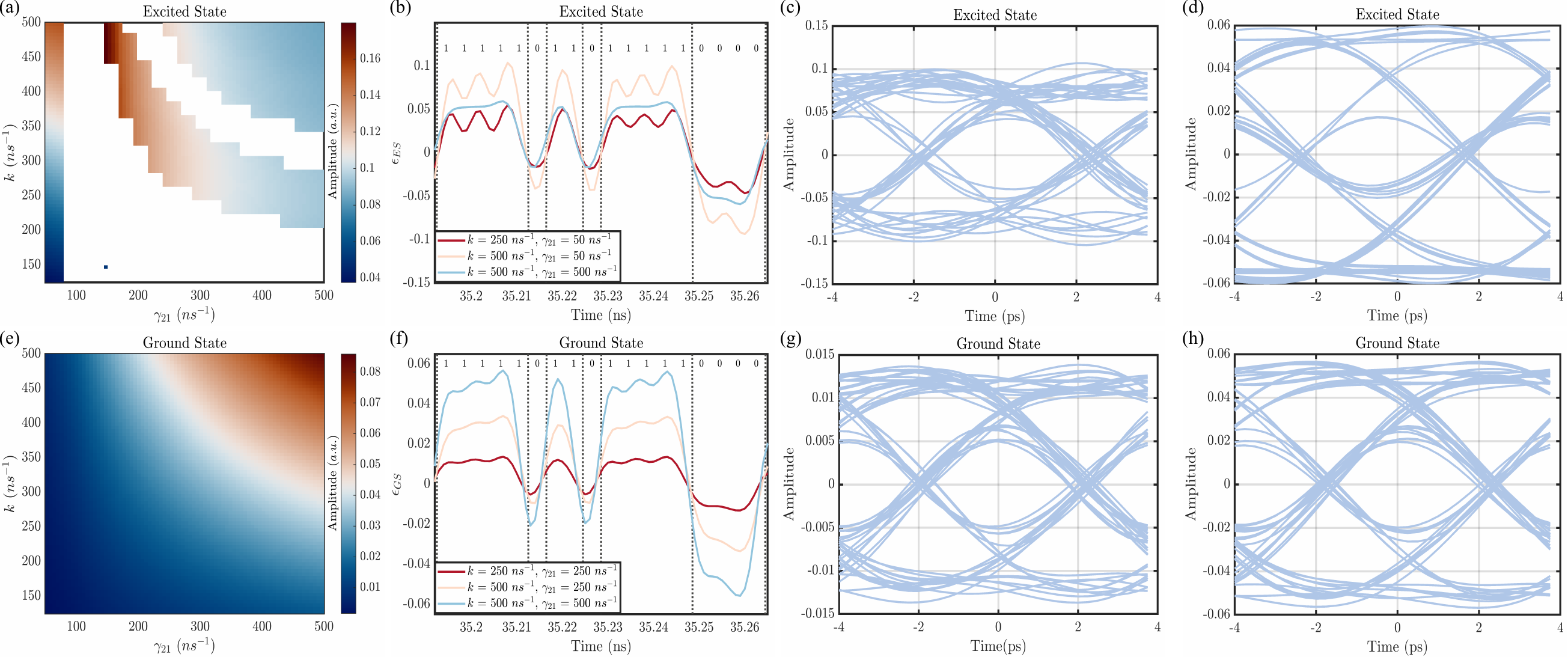}
\caption{Optical modulation amplitude maps for both ES ($a$) and GS ($e$) as a function of $\gamma_{21}$  and $k$, evolution of $\epsilon_{ES}$ $(b)$ and $\epsilon_{GS}$ $(f)$ for a  18-bit word at $250$ $Gb/s$ for three different pairs of $(k,\gamma_{21})$ and eye diagrams for ES with $k=500$ $ns^{-1}$, $\gamma_{21}$ = $50$ $ns^{-1}$ $(c)$ and $k=500$ $ns^{-1}$, $\gamma_{21}$ = $500$ $ns^{-1}$ $(d)$ and GS with $k=500$ $ns^{-1}$, $\gamma_{21}$ = $250$ $ns^{-1}$ $(g)$ and $k=500$ $ns^{-1}$, $\gamma_{21}$ = $500$ $ns^{-1}$ $(h)$.}
\label{fig:all}
\end{figure*}
Spin-lasers revealed the necessity of increasing $\gamma_{p}$ and $\gamma_{j}$ for fast polarization modulation. These can be employed for the case of QD spin-VCSELs as well. By implementing high-contrast periodic grating, large birefringence rates can be achieved for QD spin-V(E)CSEL \cite{8982841}. Despite the fact that earlier studies promoted the superiority of QD spin VCSELs due to longer spin relaxation times ($\tau_{s}$) on low temperatures or doped dots, recent results showed that for carrier lifetime of $1$ ns, $\tau_{s}$ is in the range between 5 and 8 ps for an optically pumped QD spin-V(E)CSEL lasing at room temperature \cite{doogan_evidence_2021}. 
Choosing $\gamma_{j}$ = 250 $ns^{-1}$ and $\gamma_{p}=500$ $ns^{-1}$, we demonstrate that QD spin-VCSELs can perform modulation of data for bit rates up to $250$ Gb/s (and possibly higher). This can be seen in Fig. \ref{fig:eyes} which shows the eye patterns obtained for the QD spin-VCSEL subject to polarization modulation for four different bit rates ($10$ $(a)$, $25$ $(b)$ , $100$ $(c)$ and $250$ $(d)$ Gb/s). For the two highest speed cases investigated of $100$ and $250$ Gb/s, the symmetric shape of the eye diagrams is disrupted to a certain extent and also reduced vertical eye diagram openings are obtained. In the cases of $100$ and $250$ Gb/s, although the achieved extinction ratios are reduced, these permit successful data modulation even at these ultrafast speed rates. For all cases, GS modulation efficiency is negligible with respect to the corresponding ES.

The eye diagrams indicate that QD spin-VCSELs deliver better performance to QW spin-VCSELs \cite{lindemann_ultrafast_2019} in terms of modulation efficiency. Modulation efficiency is defined as how efficiently a spin-VCSEL cavity converts a $100\%$ input circular polarized light into another output polarization. For this application, Lindemann \textit{et al.} \cite{lindemann_ultrafast_2019} demonstrated single channel ultrafast data transmission  with low amplitude of polarization oscillations, which reaches up to 1\% (0.01) for a 100\% injected spin polarization as presented on their  calculated eye diagrams. By considering the maximum value of $\epsilon_{ES}$, i.e. 5$\%$, we find that it is almost five times larger as that obtained for QW spin-VCSELs \cite{lindemann_ultrafast_2019}. We also prove that optical pumping is as suitable as hybrid pumping scheme combining electrical with circularly-polarized optical pumping for high bit rate modulation.

\section{Optimization analysis of ultrafast polarization modulation performance}

An optimization analysis of different photon lifetimes  in the cavity and intradot relaxation rates in QD spin-VCSELs can lead to even higher values of modulation efficiency for ultrafast data transmission. The investigation of the influence of the photon lifetime on the polarization oscillations is crucial since the rapid emission of the photons from cavity can prevent the mixing with photons with different ellipticities emitted after the strong subband mixing. The output ellipticity of both ES and GS is based on the interplay between fundamental physical processes; the the spin flip processes tends to equalize the gain for RCP and LCP fields, the intradot relaxation from ES to GS and the escape of carriers from GS to ES and ES to WL which couple carriers back and forth between the polarized fields. Hence, we expect that the variation of $\gamma_{21}$ will enhance the $\epsilon_{GS}$, as GS will receive carriers emitted from the pumped sublevel before electrons relax to the density with a different spin. While, the $\epsilon_{ES}$ will be decreased due to the escaped carriers from GS, which are burdened by strong subband mixing  since spin relaxation is faster than the escape procedure. ES plays an important role on GS polarization modulation. Rohm et al. \cite{rohm_ground-state_2015} and Wu et al. \cite{wu_effect_2013} demonstrated that the increase of  excited-to-ground-state relaxation time strongly limits the GS modulation bandwidth of the QD conventional laser.

The performance of QD spin-VCSELs as polarization modulation transmitter is evaluated by the optical modulation amplitude and the eye diagrams. The optical modulation amplitude is defined as the difference between two levels of eye diagrams, i.e. output ellipticity values when bits '1' and '0' are encoded. To this end, we illustrate optical modulation amplitude for both GS and ES in the plane of $\gamma_{21}$ and $k$ in Figs. \ref{fig:all} $(a)$ and $(d)$, respectively. The specific range of $\gamma_{21}$ was chosen from reported values found in QD lasers literature \cite{markus2006two,olejniczak_intrinsic_2010}. 
In the case of ES the key parameters seems to have an impact on polarization modulation performance when $\gamma_{21}$ is lower than $300$ $ns^{-1}$. The increase of $k$ increases the optical modulation amplitude as we expected, while the increase of $\gamma_{21}$ leads to lower optical modulation amplitude, a disadvantage arising from the escape of polarized carriers from GS.
The strong subband mixing yielding in an almost linearly polarized emission ($\epsilon = 0$) is inevitable in the case of GS performance since carrier recombination rate is low with respect to spin relaxation rate. This can be compensated by the increase of $\gamma_{21}$ and $k$, where the performance is benefited from the fast decay of the polarized carriers to the GS.

For high bit rate modulation it is important to  investigate the pattern dependent effects. For presentation purposes we choose to investigate the impact of three different pairs of $k$ and $\gamma_{21}$, which values provide close agreement with experimental observations, when the QD spin-VCSEL is modulated with a 18-bit word at $250$ $Gb/s$. In Fig. \ref{fig:all} $(b)$ the increase of $k$ results in higher amplitude polarization oscillation (orange curve) with a clear open eye-diagram (Fig. \ref{fig:all} $(c)$), while the increase of $\gamma_{21}$ up to $500$ $ns^{-1}$ shows strong pattern dependence and whilst this would result in a closure of the eye-diagram opening albeit
with reduced modulation efficiency (Fig. \ref{fig:all} $(d)$) yet good extinction ratio between '1' and '0' bits is achieved. In Fig. \ref{fig:all} $(f)$ it is shown that all traces have similar behaviour with increased $\epsilon_{GS}$ in the case of increasing $\gamma_{21}$ and $k$, the first pulse in a string of pulses is lower than the ones that follow it. Although both eyes for two pairs of $\gamma_{21}$ and $k$ ($k=500$ $ns^{-1}$, $\gamma_{21}$ = $250$ $ns^{-1}$ $(g)$ and $k=500$ $ns^{-1}$, $\gamma_{21}$ = $500$ $ns^{-1}$ $(h)$) are wide open.

The optimization analysis indicates that the use of $k=500$ $ns^{-1}$ and $\gamma_{21}=500$ $ns^{-1}$ provides enhancement performance, guaranteeing high polarization extinction ratio between '1' and '0' bits. The GS response in the QD spin-VCSEL shows clear open eye at $250$ $Gb/s$ data rates (Fig. \ref{fig:all} $(h)$) with modulation efficiency reaching up to $6\%$, whilst the results obtained for the QD spin-VCSEL's ES emission reveal dual ultrafast data transmission (at $250$ $Gb/s$ in Fig. \ref{fig:all} $(d)$). Therefore, importantly QD spin-VCSELs in contrast to their QW counterparts permit very high-speed data modulation from both states, offering promise for dual wavelength light sources for use as in ultrafast optical communications.

\section{Comparison of polarization modulation performance between QD and QW spin-VCSELs}

\begin{figure*}[!ht]
\centering
\includegraphics[width=\linewidth]{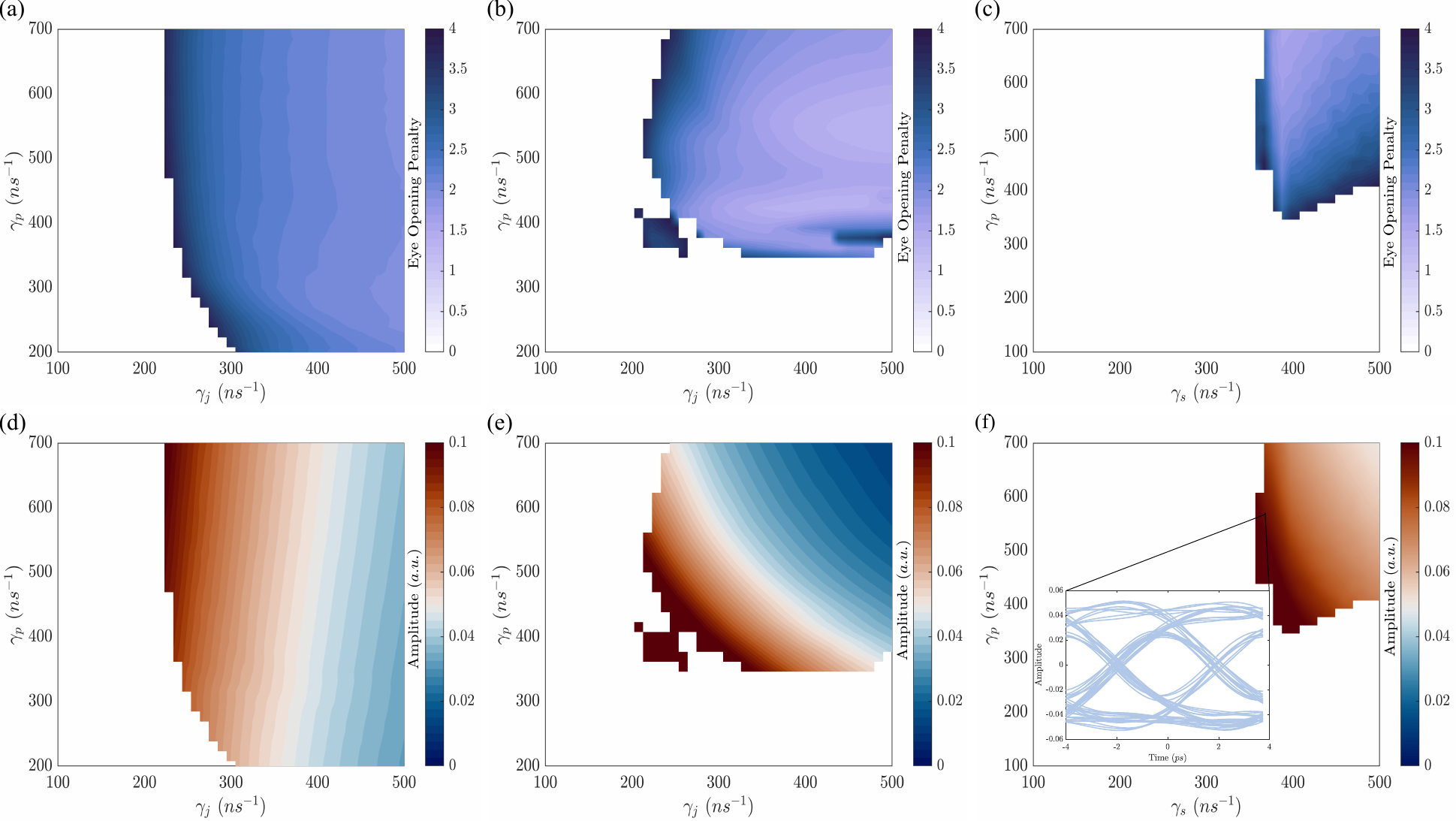}
\caption{$(a-c)$ EOP maps for GS QD spin-VCSEL, ES QD spin-VCSEL and QW spin VCSEL, respectively. $(d-f)$ The corresponding results for optical modulation amplitude. The SFM parameters are the same as in Figs. \ref{fig:all} $(d,h)$ but with $a=5$ for QW spin-VCSEL. The inset eye diagram with the maximum amplitude for $\gamma_{p}=550$ $ns^{-1}$ and $\gamma_{s}=400$ $ns^{-1}$}
\label{fig:comparison}
\end{figure*}

The superior modulation performance of QD spin-VCSELs over QW spin-VCSELs is demonstrated in this section where we compare the polarization modulation performance of GS (Figs. $\ref{fig:comparison}(a,d))$), ES (Figs. $\ref{fig:comparison}(b,e)$) and QW spin-VCSEL (Figs. $\ref{fig:comparison}(c,f)$) by means of optical modulation amplitude and EOP which is the ratio of the optical modulation amplitude and the vertical eye opening. This metric is useful for noise free systems since in binary modulated digital optical systems, waveform distortions represented by eye-opening penalty may be introduced by nonlinear effects in the transmitter and can only be manifested through time-domain measurements. We will use this here in order to find conditions resulting in suitable values of EOP where improved performance is achieved.

For the simulation of QW spin-VCSELs we employ the generalized SFM from \cite{lindemann_ultrafast_2019}. We use the same SFM parameters as in Figs. \ref{fig:all} $(d,h)$ but with $a=5$ for QW spin-VCSEL, typical value for linewidth enhancement factor in QWs. Another difference in the SFM between QW and QD spin-VCSELs is that spin relaxation rate for the case of QW spin-VCSELs  is defined as $\gamma_{s,QW}=\gamma_{n}+2\gamma_{s,QD}$ \cite{san1995light}. Here, $\gamma_{s,QW}$ is defined as $\gamma_{s}$ and $\gamma_{s,QD}$ as $\gamma_{j}$ for convenience.
The analysis of optical modulation amplitude’s maps reveal polarization oscillation with amplitude nearly equal to $0.05$ $(5\%)$ in the case of QW spin-VCSEL. 
In terms of single channel data transmission QD spin-VCSELs outperform their state-of-the-art QW counterparts in terms of modulation efficiency, yielding a two-fold improvement. Meanwhile, QD spin-VCSELs can provide two channels ultrafast data transmission with comparable modulation efficiency of each channel with QW spin VCSEL at its best conditions, increasing significantly the capacity. It is believed that the proposed high efficiency dual state QD spin-VCSEL will provide a new device oriented towards full polarization control at ultrafast regimes.

\section{Conclusion}

In conclusion, we have investigated the ultrafast polarization modulation properties of QD spin-VCSELs by means of large signal modulation. We provide a fundamental understanding of their potentials for ultrafast light sources for use in future optical communication platforms and investigate the effect of key system parameters for optimised performance.  Ultrafast bit rate polarization modulation analyzed via eye diagrams revealed modulation of data can be performed for bit rates up to $250$ $Gb/s$ (and possibly higher) with high modulation efficiencies, exceeding those achievable with QW spin-VCSELs. Notably, QD-spin-VCSELs allow high speed data modulation from both the GS and ES; hence offering great promise for ultrafast dual wavelength sources. QD spin-VCSELs offer therefore an exciting platform for ultrafast sources with optimised performance and yielding operation at hundreds of Gb/s rates for use in future optical data communication systems.

\section*{Acknowledgments}

The authors acknowledge Prof. Mike Adams of the
School of Computer Science and Electronic Engineering,
University of Essex, UK for proofreading the article and
his comments on the chosen parameters.

{\appendix[Proof of the SFM Rate Equations]
}
The complete set of Rate Equations (REs) describing the
modified SFM model reads as:
\begin{equation}\label{Appndx_f_wl}
\begin{split}
\frac{d(N_{WL}f_{WL}^\pm)}{dt}=\frac{I}{q}-\gamma_{0}N_{WL}f_{WL}^\pm(1-f_{ES}\pm)+\\\noindent{+4N_{D}\gamma_{esc}f_{ES}^\pm-\gamma_{n}N_{WL}f_{WL}^\pm\mp\gamma_{j}N_{WL}(f_{WL}^{+}-f_{WL}^{-})}
\end{split}
\end{equation}
\begin{equation}\label{Appndx_f_es}
    \begin{split}
\frac{d(4N_{D}f_{ES}^\pm)}{dt}&=\gamma_{0}N_{WL}f_{WL}^\pm(1-f_{ES}^{\pm})-4\gamma_{esc}N_{D}f_{ES}\pm-\\&-4\gamma_{n}N_{D}f_{ES}^\pm-\gamma_{21}4N_{D}f_{ES}^{\pm}(1-f_{GS}^\pm)+\\&+\gamma_{12}2N_{D}f_{GS}^\pm(1-f_{ES}^{\pm})\mp\gamma_{j}4N_{D}(f_{ES}^{+}-f_{ES}^{-})-\\&-v_{g}\Gamma{a_{ES}}{N_{D}}(2f_{ES}^\pm-1)|E_{0_{ES}}^{\pm}|^2
\end{split}   
\end{equation}
\begin{equation}\label{Appndx_f_gs}
\centering
    \begin{split}
\frac{d(2N_{D}f_{GS}^\pm)}{dt}&=\gamma_{21}4N_{D}f_{ES}^\pm(1-f_{GS}^\pm)-\gamma_{12}2N_{D}f_{GS}^\pm(1-f_{ES}^\pm)-\\&-\gamma_{n}2N_{D}f_{GS}^\pm-v_{g}\Gamma{a_{GS}}{N_{D}}(2f_{GS}^\pm-1)|E_{0_{GS}}^\pm|^2\mp\\&\mp\gamma_{j}2N_{D}(f_{GS}^{+}-f_{GS}^{-})
\end{split}    
\end{equation}
\begin{equation}\label{Appndx_f_gs}
\frac{dE_{0_{GS}}^\pm}{dt}=k[h_{1}(2f_{GS}^\pm-1)](1+i\alpha)E_{0_{GS}}^\pm-(\gamma_{a}+i\gamma_{p})E_{0_{GS}}^\mp
\end{equation}
\begin{equation}\label{Appndx_f_es}
\frac{dE_{0_{ES}}^\pm}{dt}=k[h_{2}(2f_{ES}^\pm-1)](1+i\alpha)E_{0_{ES}}^\pm-(\gamma_{a}+i\gamma_{p})E_{0_{ES}}^\mp
\end{equation}
The modified SFM REs are: 
\begin{equation}\label{Appndx_f_wl_II}
\begin{split}
\frac{df_{WL}^\pm}{dt}=\frac{I}{qA\overline{N}_{WL}}-\gamma_{0}f_{WL}^\pm(1-f_{ES}^\pm)+\frac{4N_{D}}{N_{WL}}\gamma_{esc}f_{ES}^\pm-\\
\noindent{-\gamma_{n}f_{WL}^\pm\mp\gamma_{j}(f_{WL}^{+}-f_{WL}^{-})},
\end{split}
\end{equation}
where $\overline{N}_{WL}$=${N_{WL}}/{A_{}}$.
\begin{equation}\label{n_es}
    \begin{split}
\frac{df_{ES}^\pm}{dt}&=\frac{\gamma_{0}N_{WL}}{4N_{D}}f_{WL}^\pm(1-f_{ES}^{\pm})-\gamma_{esc}f_{ES}^\pm-\\&-\gamma_{n}f_{ES}^\pm-\gamma_{21}f_{ES}^{\pm}(1-f_{GS}^\pm)+\\&+\frac{1}{2}\gamma_{12}f_{GS}^\pm(1-f_{ES}^{\pm})\mp\gamma_{j}(f_{ES}^{+}-f_{ES}^{-})-\\&-\frac{1}{4}v_{g}\Gamma{a_{ES}}(2f_{ES}^\pm-1)|E_{0_{ES}}^{\pm}|^2
\end{split}   
\end{equation}
\begin{equation}\label{n_gs}
\centering
    \begin{split}
\frac{df_{GS}^\pm}{dt}&=2\gamma_{21}f_{ES}^\pm(1-f_{GS}^\pm)-\gamma_{12}f_{GS}^\pm(1-f_{ES}^\pm)-\\&-\gamma_{n}f_{GS}^\pm-\frac{1}{2}v_{g}\Gamma{a_{GS}}(2f_{GS}^\pm-1)|E_{0_{GS}}^\pm|^2\mp\\&\mp\gamma_{j}(f_{GS}^{+}-f_{GS}^{-})
\end{split}    
\end{equation}
For the normalization of the system of REs we introduce
a change in variables as follows:
\begin{equation}\label{}
\centering
n_{WL}^\pm=\frac{N_{WL}}{N_{D}}f_{WL}^\pm
\end{equation}
\begin{equation}\label{}
\centering
E_{GS}=E_{s_{GS}}\sqrt{\frac{\gamma_{n}}{v_{g}\Gamma{a_{GS}}}}
\end{equation}
\begin{equation}\label{}
\centering
E_{ES}=E_{s_{ES}}\sqrt{\frac{\gamma_{n}}{v_{g}\Gamma{a_{ES}}}}
\end{equation}
\begin{equation}\label{}
\centering
\frac{I}{qA\overline{N}_{D}\frac{\tau_{r}}{\tau_{r}}}=\frac{I}{I_{x}\tau_{r}}=\frac{J}{\tau_{r}}=\gamma_{n}J
\end{equation}
The normalized form of modifed SFM REs (Eqs. \ref{Appndx_f_wl}-\ref{Appndx_f_es}) is:
\begin{equation}\label{n_wl}
\begin{split}
\frac{dn_{WL}^\pm}{dt}=\gamma_{n}J^{\pm}-\gamma_{0}n_{WL}^\pm(1-f_{ES}^{\pm})-\gamma_{n}n_{WL}^\pm+\\
\noindent{+4\gamma_{esc}f_{ES}^\pm\mp\gamma_{j}(n_{WL}^{+}-n_{WL}^{-})}
\end{split}
\end{equation}
\begin{equation}\label{n_es}
    \begin{split}
\frac{df_{ES}^\pm}{dt}&=\frac{1}{4}\gamma_{0}n_{WL}^\pm(1-f_{ES}^{\pm})-\gamma_{esc}f_{ES}^\pm-\gamma_{n}f_{ES}^\pm-\\&-\gamma_{21}f_{ES}^{\pm}(1-f_{GS}^\pm)+\frac{1}{2}\gamma_{12}f_{GS}^\pm(1-f_{ES}^{\pm})-\\&-\frac{1}{4}\gamma_{n}(2f_{ES}^\pm-1)|E_{s_{ES}}^{\pm}|^2\mp\gamma_{j}(f_{ES}^{+}-f_{ES}^{-})
\end{split}   
\end{equation}
\begin{equation}\label{n_gs}
\centering
    \begin{split}
\frac{df_{GS}^\pm}{dt}&=2\gamma_{21}f_{ES}^\pm(1-f_{GS}^\pm)-\gamma_{12}f_{GS}^\pm(1-f_{ES}^\pm)-\gamma_{n}f_{GS}^\pm-\\&-\frac{1}{2}\gamma_{n}(2f_{GS}^\pm-1)|E_{s_{GS}}^\pm|^2-\gamma_{j}(f_{GS}^{+}-f_{GS}^{-})
\end{split}    
\end{equation}
\begin{equation}\label{E_s_GS}
\frac{dE_{GS}^\pm}{dt}=k[h_{1}(2f_{GS}^\pm-1)](1+i\alpha)E_{GS}^\pm-(\gamma_{a}+i\gamma_{p})E_{GS}^\mp
\end{equation}
\begin{equation}\label{E_s_ES}
\frac{dE_{ES}^\pm}{dt}=k[h_{2}(2f_{ES}^\pm-1)](1+i\alpha)E_{ES}^\pm-(\gamma_{a}+i\gamma_{p})E_{ES}^\mp
\end{equation}
{\appendix[Indicative Results for $\gamma_{\alpha}$ and $J$]
In this Appendix we focus on the effect of the parameters on the polarization oscillation amplitude. We simulate parameter sweeps for a single parameter at a time, keeping all others at the values given in Figs. \ref{fig:all} $(d)$ and $(h)$. 

The increase of $J$ from $100$ to $200$ almost doubles the polarization oscillation amplitude for both GS and ES. Hence, $J$ is a crucial parameter toward higher oscillation amplitudes but its increase can cause unwanted heating in the device. Dichroism rate $\gamma_{a}$ is linked to the magnitude of birefringence of the cavity and we sweep this parameter in a large region $(-4 ns^{-1} - 4 ns^{-1})$, but its effect is negligible. So, we choose $\gamma_{a}=0$ for our simulations. 
\begin{figure}[h]
\centering
\includegraphics[width=\columnwidth]{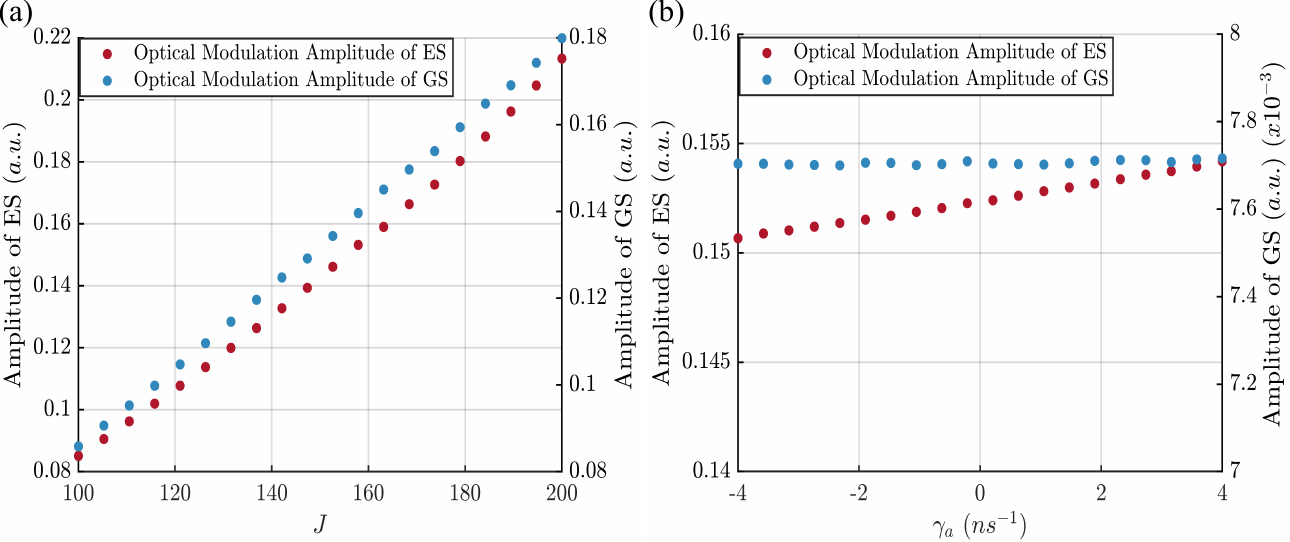}
\caption{Optical modulation amplitude for both GS (right column) and ES (left column) as a function of $J$ $(a)$ and $\gamma_{a}$ $(b)$.}
\label{fig:appndx}
\end{figure}

}


%

\bibliographystyle{IEEEtran}
\bibliography{IEEEabrv,sample}

\vspace{-4mm}\begin{IEEEbiographynophoto}{Christos Tselios}
received his bachelor’s degree in Physics from the University of Patras in 2019. Between 2019 and 2021, he studied at the Department of Materials Science of the University of Patras where he received the Master’s degree in Applied Optoelectronics. He is currently pursuing his doctoral degree in Department of Electrical and Computer Engineering of the University of Peloponnese. Since February 2022, he is also a member of the Industrial Systems Institute of Athena Research Center. His main research interests are focused on ultrafast signal transmission in optical communication systems and simulation of optoelectronic devices.
\end{IEEEbiographynophoto}

\vspace{-4mm}\begin{IEEEbiographynophoto}{Panagiotis Georgiou} received the B.Sc Degree in Materials Science from the University of Patras, Patras,Greece, in 2017 and the M.Sc degree in Materials Science from the same institute in 2020. Since 2020 he is working towards his PhD degree in Optoelectronics at the Department of Materials Science, University of Patras, Greece. His research interests include Nonlinear Dynamics in Quantum Dot spin-VCSELs, Photonic Reservoir Computing and Photonic Spiking Neural Networks.
\end{IEEEbiographynophoto}

\vspace{-4mm}\begin{IEEEbiographynophoto}{Christina (Tanya) Politi}
received a B.Sc. in physics from the University of Athens in 1998 and a M.Sc. degree in the "Physics of laser communications" from the University of Essex in 2000. Subsequently, she joined the Photonic Network Research Group in the Department of Electronic Systems Engineering at the University of Essex where she worked until early 2005 and obtained her PhD. In 2005 she joined the National Technical University of Athens – Telecommunications Laboratory as a research associate where she worked in various IST and national projects. During this period she was also a visiting lecturer at the University of Peloponnese, which she joined at the end of 2008 as an Assistant Professor.  Currently, she is Assistant Professor at the Department of Electrical and Computer Engineering , University of Peloponnese. She has a long experience in designing and providing optoelectronic components, systems and networks for optical communications as well as in the area of novel application paradigms and numerous collaboration with Greek and European organizations. She is a co-author of over 100 international journal and conference publications in the area of communication systems and subsystems and networks. Her research interests include optical networking and optical switching with emphasis on optical switch architectures and subsystems, optical packet and circuit switched networks, high speed optical networks and data center interconnection. Recently her work focuses on innovative technologies in digital infrastructures for IoT applications, high performance wireless and wired technologies for application areas like smart cities and future hospitals.
\end{IEEEbiographynophoto}

\vspace{-4mm}\begin{IEEEbiographynophoto}{Antonio Hurtado}
received the Ph.D. degree from the Universidad Politécnica de Madrid (UPM), Madrid, Spain, in December 2006. He is currently a Reader and Turing Artificial Intelligence (AI) Fellow with the Institute of Photonics, University of Strathclyde, Glasgow, U.K. He has more than 15 years’ of International Research Experience in photonics, Universities of Essex and Strathclyde, U.K., University of New Mexico, USA, and UPM, Spain. His research interests include neuromorphic photonics, laser nonlinear dynamics, nanolaser systems, and hybrid nanofabrication. He was the recipient of two Marie Curie Fellowships by the European Commission: Projects ISLAS (2009–2011) and NINFA (2011–2014), a Chancellor’s Fellowship by the University of Strathclyde following which he was appointed as a Lecturer with the Institute of Photonics, Department of Physics, University of Strathclyde, in 2014, and a Turing AI Acceleration Fellowship by the U.K. Research and Innovation Office and the U.K. Government Business, Energy and Industrial Strategy Department to develop a five-year Research Programme on Photonics for Ultrafast AI, in 2020.
\end{IEEEbiographynophoto}

\vspace{-4mm}\begin{IEEEbiographynophoto}{Dimitris Alexandropoulos}received the B.Sc degree in physics in 1998 from the University of Athens, Athens, Greece, and the M.Sc. degree in the physics of laser communications in 1999 from the University of Essex, Colchester, Essex, U.K., and the PhD degree for work on GaInNAs-based optoelectronic devices, from the same University in 2003. After one year of postdoctoral work at the University of Essex he moved in 2004 to the Optical Communications Laboratory, Department of Informatics and Telecommunications, University of Athens. He is now an Assistant Professor  in nanophotonics at the Department of Materials Science, University of Patras, Patras, Greece. He is also a collaborating academic faculty member of the Industrial Systems Institute, Athena Research Center. He has co-authored more than 80 journal and conference publications. His research interests include novel photonic materials, design and fabrication of holographic optical elements, photonic sensors, integrated photonics and optoelectronic devices.
\end{IEEEbiographynophoto}

\newpage

\vspace{11pt}

\vfill

\end{document}